# SLM-based Digital Adaptive Coronagraphy: Current Status and Capabilities

Jonas Kühn,*[a] Polychronis Patapis,[a] Xin Lu,[a] and Marcel Arikan[a]

[a]Institute of Particle Physics and Astrophysics, ETH Zurich, Wolfgang-Pauli-Strasse 27, 8093 Zurich, Switzerland

**ABSTRACT**

Active coronagraphy is deemed to play a key role for the next generation of high-contrast instruments, notably in order to deal with large segmented mirrors that might exhibit time-dependent pupil merit function, caused by missing or defective segments. To this purpose, we recently introduced a new technological framework called digital adaptive coronagraphy (DAC), making use of liquid-crystal spatial light modulators (SLMs) display panels operating as active focal-plane phase mask coronagraphs. Here, we first review the latest contrast performance, measured in laboratory conditions with monochromatic visible light, and describe a few potential pathways to improve SLM coronagraphic nulling in the future. We then unveil a few unique capabilities of SLM-based DAC that were recently, or are currently in the process of being, demonstrated in our laboratory, including NCPA wavefront sensing, aperture-matched adaptive phase masks, coronagraphic nulling of multiple star systems, and coherent differential imaging (CDI).

## 1. INTRODUCTION

In the past two decades, new exoplanet detections have accumulated by the thousands, essentially through the use of radial velocity (RV) and transit techniques. In parallel, direct "high-contrast" imaging (HCI) has a synergistic potential for observing exoplanets, as not only does it allow in-situ observing in the host stellar environment (e.g. to study planet formation and disk interaction processes), but it also alleviates part of the observational biases from RV and transit in favor of massive close-in objects, and opens the door to spectroscopic characterization of exoplanetary atmospheres. Yet, since the initial promising discoveries of Fomalhaut b[1] and HR8799 b,c,d,e[2] less than 10 years ago, the number of newly imaged planetary mass objects with HCI has not exceeded a dozen. This is in spite of the recent commissioning of several second-generation so-called "planet-imager" instruments, including the Gemini Planet Imager (GPI),[3] the Spectro-Polarimetric High contrast Exoplanet Research (SPHERE) instrument,[4] and the Subaru Extreme-Adaptive Optics (SCExAO) facility.[5,6] These near-infrared (NIR) HCI instruments incorporate the latest Adaptive Optics (AO) systems to correct for the atmospheric turbulences, to reach the extreme AO regime with Strehl ratio in excess of 0.8. Furthermore, these facilities make use of the latest techniques of wavefront sensing (WFS) to correct for non-common path aberrations (NCPAs) present in the beam train between the AO WFS and the science focal-plane array (FPA),[7,8] and integrate coronagraphs to mask, or diffract away, the bulk of the on-axis stellar point-spread function (PSF). Finally, those instruments rely on advanced observational strategies, among which angular differential imaging (ADI), spectral/dual-band differential imaging (SDI/DBI), polarimetric differential imaging (PDI), etc. However, the outcome in terms of new directly-imaged exoplanets has been admittedly modest so far (a dozen confirmed detections), although the yield in debris or protoplanetary disk imaging has been considerable. This is partially because the above mentioned coronagraphy techniques and observational strategies tend to lose effectiveness at close angular separation, where it matters the most.

*jonas.kuehn@phys.ethz.ch

Indeed, detecting exoplanets with HCI from the ground is admittedly a daunting task, due to the extreme contrast ratio ($10^{-6}$ to $10^{-12}$) required at angular separations as small as few diffraction beam widths. The upcoming class of extremely large telescopes (ELTs) will ease some of these challenges, notably by improving sensitivity and angular resolution, but other issues affecting direct imaging and coronagraphy will remain, or may even worsen. Those include residual fast atmospheric residuals, tip/tilt jitter, differential atmospheric refraction, resolved nearby stars, NCPAs, pupil registration stability, and non-ideal segmented telescope apertures. Some of these aspects are inherent to the larger ELTs apertures and support structures, but notably also arise from the segmented geometry of the ELTs primary mirrors, whose merit functions may evolve over time due to dead, defective or missing individual mirror segments. The atmospheric wavefront correction requirements are obviously less of a concern in space, although the post-Hubble era will also see the appearance of segmented primary mirrors, and possibly similarly related issues with even less margin for in-situ intervention.

## 2. SLM-BASED ACTIVE CORONAGRAPHY

### 2.1 Rationale for active coronagraphy concepts

The idea that deeper integration of adaptive optics components within future high-contrast instruments should include some sort of active coronagraph was recently discussed by a relevant fraction of the HCI research community, during the "Optimal Optical Coronagraphs" workshop. This meeting took place at the Lorentz Center in Leiden, The Netherlands, on September 25-29, 2017, and several key arguments for future active coronagraphy concepts (in amplitude and/or phase) were put forward at this occasion. The outcome of the discussions that occurred during the splinter session dedicated to active coronagraphy are summarized here below, while the reader also interested in the other topics reviewed during this workshop should refer to the corresponding summary papers.[9-11]

In essence, three different – yet complementary – usage scenarios of active dynamically-reprogrammable coronagraphs can be differentiated, to tackle various challenges:

- Environmental constrains:

    This primarily includes dealing with issues originating from segmented and non-static telescope pupils (time-evolving pupil figure, reflectivity non-uniformity for each mirror segment, missing/defective/dead segments), but also residual tip/tilt jitter, wind effects (AO "butterfly"), thermal breathing, residual atmospheric dispersion, chromatic and polarization effects, and available instrument volume envelope. The latter is particularly relevant to space-based observatories, where less optics and less mechanical actuators are generally preferred, but it also matter for mid-IR ground-based instrumentation (as less components generally means less thermal background).

- Science case constrains:

    Those are numerous and sometime hardly predictable in advance, but there is a clear on-going trend towards two kind of operating modes for an HCI instrument and its – ideally reconfigurable - coronagraphic stage: discovery (blind search in "survey" mode), and characterization (i.e. follow-ups, where the location of companion can be estimated from a previous observation, notably from RV epochs). Other observationally-challenging science cases include multiple stars systems or partially resolved giant stars (a foreseeable issue in the ELTs era), where some kind of ad-hoc coronagraphic configuration would also be desirable.

- Laboratory developments:

    Active coronagraphs could fill an important niche to test new phase or amplitude patterns in the lab, ahead of manufacturing. And this not only to prototype a coronagraphic mask, but also to implement an active reconfigurable pupil-plane apodizer, upstream or downstream of a focal-plane mask.

We propose to define such active optics or adaptive coronagraph as a component whose properties can be actively modified to increase the signal-to-noise (S/N) ratio of an astrophysical source. It would in consequence require some kind of feedback signal, while it conversely could – and this is a key argument – also be used to probe the scientific field-of-view, as a way to perform coherence differential imaging (CDI).[12] One can therefore expect developing some flavor of a close-loop system in the longer run, similar to an AO stage. This would translate into transforming the system consisting of the telescope AO facility and the HCI instrument into an "integrated AO system", a likely evolution that would optimally integrate newly developed data processing algorithms and real-time computing technologies.

An important milestone for the community is the work undertaken by various research groups to optimize the putative coronagraph of the WFIRST-AFTA telescope and its complex pupil aperture, making use of a pair of active DMs to dynamically apodize the wavefront in both amplitude and phase, and in presence of tip/jitter and slowly-evolving NCPAs.[13,14] In some sense, one can also view all the works on "dark hole" generation using pupil-plane DMs to actively dig high-contrast regions in the focal-plane as some sort of active pupil-plane coronagraphy.[15] Although we have been alone in investigating focal-plane phase coronagraphy using liquid-crystal (LC) SLMs so far, some other research groups are looking into LC-SLMs as pupil-plane modulation devices for HCI,[16] or as "light valves" for focal-plane vortex coronagraphy of multiple stars systems.[17] We can also cite the inspiring pioneering work in Bourget et al. 2012,[18] where they tried to use LC bubbles to as tunable amplitude or phase coronagraphs

## 2.2 Spatial-light modulators (SLM) as focal-plane phase coronagraphs

It is beyond the purpose of this section to describe the pixelated LC-SLM technology in details, as this has been already extensively covered elsewhere.[19-21] Of main interest for high-contrast applications is the liquid crystal on-silicon (LCOS) sub-category,[22] which can operate in reflection configuration with very compelling performance metrics in terms of fill-factor, pixel pitch, modulation dynamic range, and framerate. The main working principle of a parallel-aligned nematic (PAN) LCOS SLM panel arises from a linearly-polarized impending beam seeing variable birefringence depending on the amount of tilt of the LC molecules, which is a function of the induced pixel-scale electric field (see Figure 1 in Kühn et al. 2016).[23] Using tunable birefringence to change the refractive index of the LC material, one can in turn set the optical path length (OPL), i.e. the phase delay, for each individual pixel. Most commercially-available LCOS panels are sold by three companies, Holoeye (DE), Meadowlark (US) and Hamamatsu (JP), with the main markets being telecom fiber communication (optical switch) and projector displays.

Overall, LCOS SLMs provides exquisite spatial sampling (>90% fill factor, <10 μm pixel pitch, resolution of several millions pixels), a key argument for focal-plane applications where the stellar PSF has to be sufficiently well sampled to be able to program complex phase maps, combined with decent phase modulation "stroke" and operational wavelength range (up to one wave retardance at 1.55 μm, possibly at 2.2 μm). Further, although most commercially-available panels operate around 30 to 60 Hz refresh rate (video-rate), we note that Meadowlark recently released "high-speed" SLMs operating at up to 500-700 Hz: this kind of temporal bandwidth starts to reach interesting levels, with refresh timescales in the order of the typical atmospheric coherence time, at least in the NIR. As for spatial sampling, Holoeye also recently introduced a new "Gaea" 4K LCOS panel with a pixel pitch of ~3.5 μm: this has direct implications for the spatial envelope of a putative real telescope instrument, as it would, for example, translate into a spatial sampling of 10 pixels per λ/D focal-plane resolution element, under a very realistic focal ratio of F/25 at H-band. This supremely small pixel pitch range unambiguously represents a large part of the current LCOS-SLM technological advantage over classical DMs, with larger pixel pitch in the order of hundreds to thousands μm.

There are however a few numbers of potential shortcomings or hurdles to deal with when using LCOS SLMs, which might limit the applicability of the technology for astronomical HCI, at least at certain wavelengths regimes, or complicate its implementation into a telescope instrumentation environment. First, the intrinsic throughput is far from optimal (~60%, the rest being either absorbed or diffracted into higher order modes), and those panels require linearly polarized light as an input. Hence at least 50% of the incoming non-polarized starlight has to be either thrown away, or redirected towards another optical arm (potentially also including a SLM-based active coronagraph), which considerably complicates either the sensitivity, or the optical design and spatial envelope of an instrument. Second, the SLM-induced phase shift is highly chromatic, as it relies on a scalar phase delay instead of a geometric phase shift like, e.g., the vector vortex[24] or vAPP[25] coronagraphs. One can estimate the related theoretical chromatic coronagraphic null depth leakage from Riaud et al. 2003[26] as about $2 \cdot 10^{-3}$ for 10% bandwidth at H-band: this is a rather serious limiting effect, albeit still negligible by roughly an order of magnitude as compared to routinely achieved on-sky null depths with more achromatic coronagraphs,[6] where wavefront control errors are still dominating the leakage budget. However, as ground-based AO performance and low-order NCPAs (especially tip/tilt jitter control) correction progress in future instruments, one can expect this limitation to detrimentally kick in and set a hard contrast limit, and even more immediately so for space-applications or broadband implementations (as typically required for an integral field spectrograph unit). There are nevertheless several avenues for mitigating this chromatic leakage that have been proposed in the past, which are worth exploring and might end up being particularly suitable for an active coronagraph. Among them is the Roddier&Roddier coronagraph,[27] which relies on several concentric rings with various phase shifts in the focal-plane to alleviate the chromaticity of the scalar phase delay at the center. A similar approach is being actively tested on the SCExAO instrument to improve the bandwidth of the phase-induced amplitude apodization (PIAA) coronagraph.[28] Third and last, active SLMs operating temperature requirements are typically in the range of 10 to 40°C (source: Meadowlark), which make them unsuitable for integration in a cryogenic environment like required for most thermal-IR wavelengths instruments (e.g. L-N bands imaging, i.e. 3.5 – 10 µm). In addition, the L-band (3.5 µm) usually corresponds to a strong absorption band of LC material, further disqualifying this technology for mid-IR instrumentation, where another approach for active coronagraphy will be required.

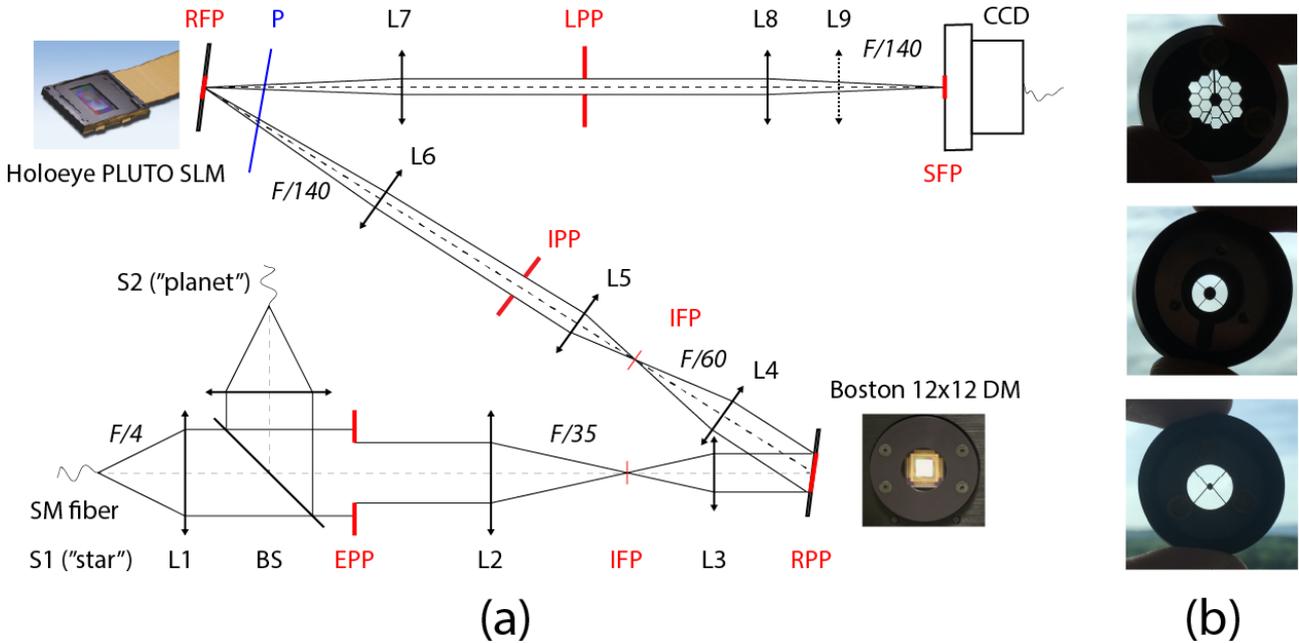

Figure 1. (a) Optical layout of the reflective-SLM DAC high-contrast imaging testbed (not to scale, reflection angles are exaggerated). S: unresolved point sources; L: aspheric lenses; BS: beam-splitter; (E/R/I/L)PP: Entrance/Reflective/Intermediate/Lyot pupil planes; (I/R/S)FP: Intermediate/Reflective (coronagraphic)/Science focal planes, P: linear polarizer. L8 and L9 are foldable lenses to enable focal-plane, respectively pupil-plane imaging.
(b) Examples of laser-cut Molybdenum pupil masks manufactured in-house, to be inserted at various pupil-plane locations.

## 2.3 SLM-based active coronagraphy: Early results

In order to evaluate the potential of using LCOS reflective SLMs as programmable active focal-plane phase mask (FPM) coronagraphs, we started a dedicated research effort back in late 2016. A monochromatic high-contrast optical testbench was built in our laboratory, relying on aspheric relay lenses to achieve diffraction-limited performance at the operating wavelength (633 nm). As shown on Fig. 1a, a single-mode fiber is used as a proxy for an unresolved point source (the star), and the beam is then collimated towards an aperture plane that defines an entrance pupil plane (EPP, aka the telescope). Various such telescope pupil aperture masks can be inserted at the EPP location, and are manufactured in-house by laser cutting 0.1-mm thick Molybdenum (see Fig.1b). To be able to perform some experiments with two point sources, for example to simulate a binary star or a star & planet configuration, a thin plate beam-splitter (BS) is inserted upstream of the EPP, effectively enabling to insert a second (temporally incoherent) collimated beam. The BS can then be adjusted in tip and tilt to "move around" the second point source in all subsequent focal planes. Downstream of the EPP, the beam is relayed towards a reflective intermediate pupil-plane (RPP), where a 12x12 deformable mirror (DM) from Boston Micromachines is located. The DM is used to correct for low-order aberrations on the bench (see §3.2), and to be able to artificially insert coherent speckles in the field-of-view. Then the beam is slowed down to a focal-ratio of ~ F/140, in order to achieve a spatial sampling of at least 10 SLM pixels per λ/D units in the coronagraphic reflective focal-plane (RFP), where the LCOS SLM is located (Holoeye PLUTO-VIS-014). The off-axis angle of this optical arrangement is kept at a minimum (here 3.5 deg) to prevent crosstalk effects between the SLM pixels, and a linear polarizer (P) is placed in front of the SLM panel, simultaneously ensuring that only linearly-polarized light can reach and exit the phase modulator. Finally the beam is relayed towards a subsequent Lyot pupil-plane (LPP), where appropriately scaled Lyot masks can be inserted, in a similar fashion as in the EPP, to mask out coronagraphically-diffracted on-axis light, before being focused towards the science focal-plane (SFP) on a CCD camera (pco.pixelfly). Further details on the optical layout, and SLM panel specifications and calibration, are provided in Kühn et al. 2017.[29]

An attractive consequence of using an active SLM display is the possibility of programming various focal-plane phase patterns as a software-only operation. This is illustrated in Fig. 2a,, where a variety of coronagraphic phase masks currently in use on real-world HCI instrument are sequentially programmed onto the SLM: a four-quadrant phase mask (FQPM), an eight-octant phase mask (EOPM), and a vortex phase mask[24] of topographic charge 2 and 4. The resulting intensity distribution in the downstream Lyot pupil-plane for each of those FPMs is also shown on Fig. 2a. Being able to instantaneously switch from one FPM coronagraphic phase pattern to another would represent a compelling advantage on-sky, for example to adapt to observational conditions (e.g. increase a vortex topographic charge when in presence of stronger tip/tilt jitter). An example of measured raw contrast curves for the non-coronagraphic and coronagraphic (vortex charge-2) cases is depicted in Fig. 2b. The typical measured raw coronagraphic null depth achievable with our SLM panel is in the order of $1.5 \cdot 10^{-2}$, when correcting for low-order aberrations with the DM (see §3.1)

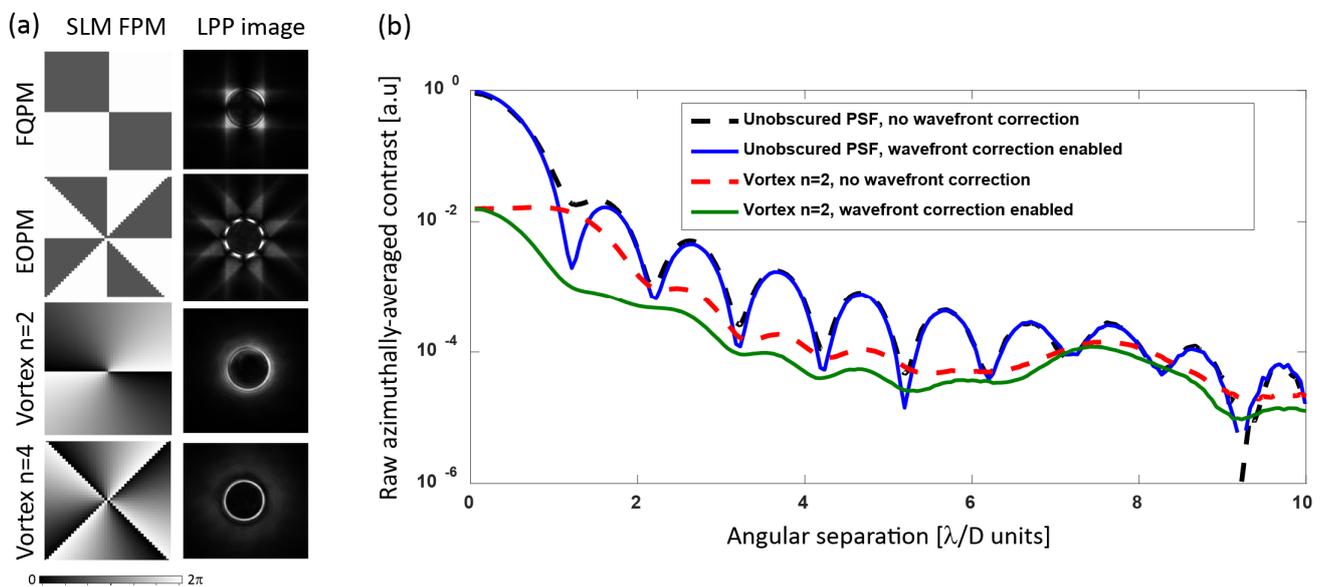

Figure 2. (a) Example of coronagraphic focal-plane phase masks (FPMs) program that can be programmed onto the SLM, and corresponding Lyot pupil-plane (LPP) recorded intensities. (b) Example of achievable raw contrast curve for the vortex of topographic charge n=2, with or without correction of low-order aberrations with the DM (best null depth $N \sim 1.5 \cdot 10^{-2}$).

In an earlier research phase back in 2015-16, we also investigated the potential of transmissive twisted nematic (TN) LC-SLMs as programmable FPMs that could be directly inserted in the beam train of an existing instrument,[23] as opposed to the unusual off-axis reflective arrangement required to operate reflective PAN LCOS SLMs (Fig. 1a). Using an Holoeye LC-2012 transmissive SLM, we were able to achieve coronagraphic null depths at low as $\sim 7 \cdot 10^{-3}$, albeit suffering from a higher speckles floor of around $10^{-4}$, which was likely caused by higher inter-pixel anisotropic response (in amplitude and phase). Additionally, these TN SLMs suffer from a cross-talk between phase and amplitude modulation (i.e. the response is not "phase-only", but "phase-mostly"), exhibit a very low throughput (< 0.1) as the fill factor is low (< 0.6) and they have to be operated between two linear polarizers, and the pixel pitch is very large (~ 35 µm), which requires very slow optics. Astronomical applications are further impeded by the low dynamic range of the phase modulation, which barely reach π at 633 nm, hence preventing any realistic implementation inside a near-infrared (NIR) instrument. Further details on these preliminary adaptive coronagraphy experiments with a transmissive TN SLM are provided in Kühn et al. 2016.[23]

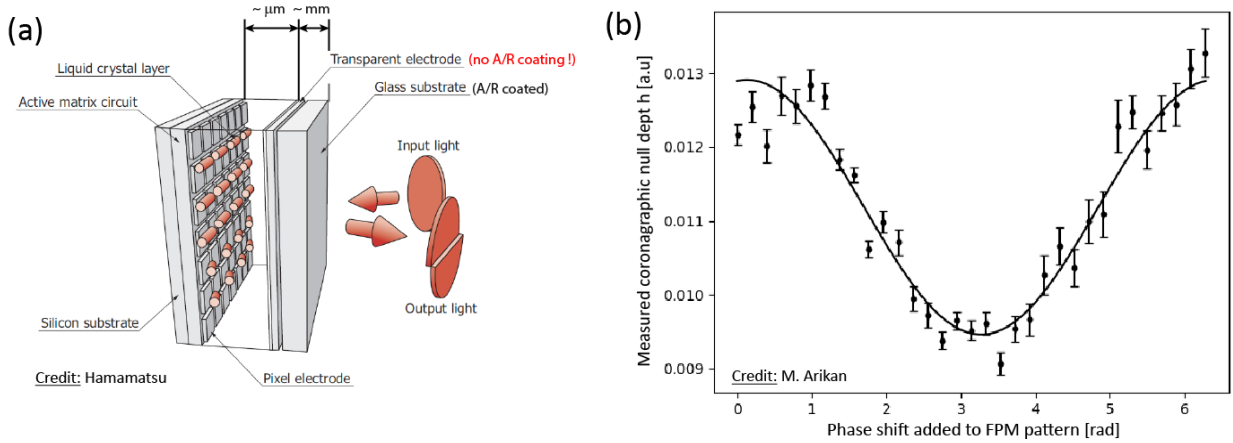

Figure 3. (a) Cartoon of the internal geometry and active layers of a phase-only LCOS SLM, with the LC layer thickness grossly exaggerated (a few μm optical thickness) as compared to the quartz front glass window (a few mm thickness). (b) Measured coronagraphic null depth (vortex n=2) in function of applied phase shift to the coronagraphic phase map.

## 2.4 LCOS-SLMs as FPM coronagraphs for HCI: Performance limitations and potential workarounds

As can be seen in Fig.2b, routinely achievable coronagraphic null depth with LCOS reflective SLMs – and in monochromatic conditions - is in the order of only $1.5 \cdot 10^{-2}$, at least with our Holoeye PLUTO device. The fact that transmissive TN SLMs with admittedly worse overall specifications can reach twice better nulls (see §2.3 and Kühn et al. 2016)[23] suggests that the reflective geometry of the beam path might play an important role as a leakage source. This effect is frequently referred as "zero order term" in the SLM literature,[30] as most SLM applications involve some kind of diffractive setup where a grating function is programmed onto the SLM display. As a first step, we did slightly modify the optical layout of Fig.2a, by aligning the SLM panel perpendicular to the incoming beam and adding a grating function on top of our coronagraphic FPM phase patterns. A quick investigation showed that deeper null depths of a few $10^{-3}$ could be reached this way (data not shown), although relying on a diffraction grating is inherently impractical with broadband light in the longer term, as it would convert wavelength into spatial shear. Nevertheless, these results hints at on-axis leakage terms unrelated to the phase modulation effect, instead likely linked to parasitic reflections at the various air-glass-LC optical interfaces within the LCOS SLM substrate structure (see Fig.3a). As indicated on Fig.3a, although the front air-glass interface is anti-reflection (A/R) coated, with the manufacturer quoting a reflectivity of less than 0.5%, the glass-LC interface with the transparent electrode is not. This is obviously due to the changing nature of the LC layer refractive index when the SLM is in operation, preventing to design an optimal A/R coating at this location, and the manufacturer indicates typical reflectivity at this interface in the order of 2% (source: Holoeye), in good agreement with the measured null depth of Fig.2b. Both to verify and mitigate this effect, we implemented the idea of Zhang et al. 2009,[30] who suggest to phase-shift the phase pattern programmed onto the SLM to achieve a destructive interference with the zero-order leakage term. Indeed, as shown in Fig.3a, given that the thickness of the LC layer is only a few tens of μm while the glass window is a few mm optically thick, and that we employ a 633-nm laser diode with limited coherence length, we can expect the LC-modulated reflected beam to solely interfere with the suspected parasitic reflection. Figure 3b presents the results of phase-shifting the programmed coronagraphic FPM phase pattern, by displaying the null depth in function of the applied phase offset. As can be seen, a clear modulation signature is present, and null depths slightly better than $10^{-2}$ can be achieved this way, if applying a phase shift around π. This level of null depth is still very modest, but we note that it is still very rarely achieved on-sky, even with latest extreme AO instruments.[6]

Further custom optimization steps to mitigate the parasitic reflection at the glass-LC interface would involve operating with a thicker high-retardance SLM panel, where the relative change of refractive index of the LC material is smaller, hence possibly opening the door to designing some kind of A/R coating for this optical interface. Even more so, we note that the manufacturer quotes a phase flicker standard deviation of $\sigma_\phi \sim 0.2$ rad for our PLUTO-014 panel, which theoretically could limit the null depth to $\sim 1.2 \cdot 10^{-2}$,[26] while higher retardance panels typically exhibiting phase flicker specifications nearly ten times lower. We therefore intend to test high retardance panels in the future, also from other manufacturers, which are typically slower (SLM response time goes from 15 to 60 ms or more), but might deliver better null depth performance. In this regards, we also recently initiated efforts to characterize SLMs panels at the individual pixel scale, to better evaluate phase precision and stability, and inter-pixel (cross-talk) effects.

## 3. ADAPTIVE CORONAGRAPHY: A FEW APPLICATION EXAMPLES

### 3.1 A self-calibrating coronagraph based on the phase-shifting Zernike wavefront sensor

One main contrast limitation in high-contrast imaging arise for so-called non-common path aberrations (NCPAs) occurring downstream of the AO wavefront sensor (WFS), along the scientific beam path. The resulting NCPA speckles in the final science focal-plane – which are often called "semi-static" as they evolve within time frames of seconds to hundreds of seconds due to thermally- and mechanically-induced flexure - limit achievable contrast, and are a frequent source of false positive detections. As mentioned in §1, various observational strategies (ADI, SDI, PDI) to estimate the stellar PSF and subtract it from the reduced science product can be employed, but all these schemes have in common that they tend to increasingly lose effectiveness at close angular separation. To address this issue, various WFS have been proposed or implemented to measure NCPAs along the scientific beam train,[11] either in an interleaved fashion before or in-between science observations, or in real-time during the scientific exposures. The wavefront correction is then generally achieved by adding centroid offsets to the AO DM solution. Among those NCPAs WFS techniques is the Zernike wavefront sensor (ZWFS), successfully implemented as the "ZELDA" module on the SPHERE HCI instrument.[8] The original idea arise from the Zernike phase-contrast microscopy technique, where the central core of the PSF is phase-shifted in some intermediate focal-plane, subsequently converting phase aberrations (or phase-only objects like cellular material) into detectable intensity signal downstream. In the ZELDA scheme,[8] the phase-shift $\Delta\phi$ is unique and set to $\Delta\phi = \pi/2$, which is in practice achieved by inserting a dedicated phase-mask in an intermediate focal-plane, typically in place of a coronagraphic FPM, and the wavefront measurement is done in the pupil-plane. Another more comprehensive, albeit less straightforward to implement, way to run the ZWFS scheme has been previously proposed by Wallace et al. back in 2011: the phase-shifting Zernike wavefront sensor (PS-ZWFS).[7] Here four sequential phase shifts are applied to the core of the PSF, resulting in the phase-diversity pupil-plane complex wavefronts $E_n$ described by Eq.1:

$$E_n \sim A(e^{i\Delta\varphi_n} + \varepsilon + i\phi) \qquad (1)$$

where $Ae^{i\phi}$ is the input wavefront free of NCPAs (unaberrated), $\Delta\varphi_n$ is the applied phase-shift to the PSF core in the intermediate focal-plane (typically $\Delta\varphi_n = -\pi/2 + (n-1)\cdot\pi/2$), $\varepsilon$ corresponds to amplitude aberrations, and $\phi$ to phase aberrations ($\phi \ll \pi$). Then, following the classical phase-shifting interferometry formalism, one can show that phase and amplitude aberrations can be estimated from Eq.2:[7]

$$\begin{cases} \phi \sim \frac{I_3 - I_1}{I_0} \\ \varepsilon \sim \frac{I_2 - I_4}{I_0} \end{cases} \qquad (2)$$

with $I_n = E_n \cdot E_n^*$ and $I_0 = (I_1 + I_2 + I_3 + I_4)/4$. This PS-ZWFS scheme therefore enables to retrieve both phase and amplitude NCPAs from four successive pupil-plane measurements, with improved static noise rejection as compared to the single phase shift ZWFS measurement, given the differential nature of the estimation. However, in practice it does require either 3-4 different phase masks, or some kind of focal-plane mechanical phase shifting device, as original proposed.[7]

Taking advantage of the versatile nature of our programmable SLM active FPM coronagraphs, putting into practice the PS-ZWFS scheme is however as simple as a software-only sequential change of phase pattern in the reflective intermediate coronagraphic focal-plane. No mechanical actuation is required at this level, although pupil-plane imaging capability is required (in our case, a lens has to flipped in front of the camera in Fig.1a). Figure 4 presents our retained PS-WZFS implementation approach (Kühn et al. in prep) and actual phase NCPAs measurement, and subsequent correction with our 12x12 DM after spatial-filtering and multiplication with the inversed interaction matrix (built with 28 Zernike modes).

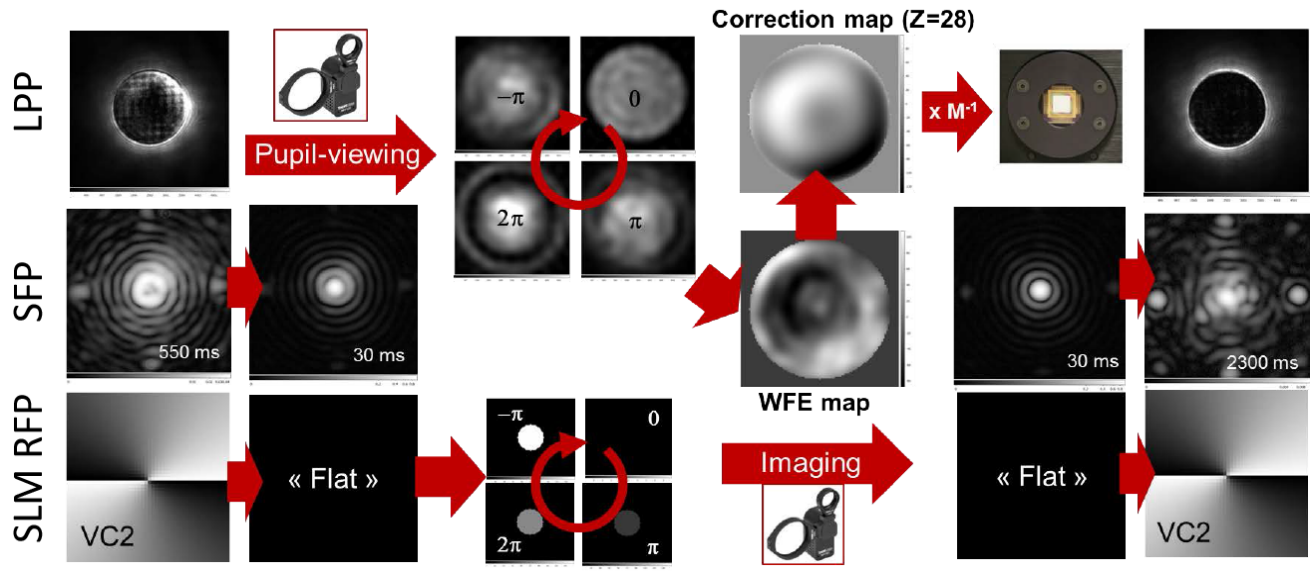

Figure 4. Operational flow chart of the "self-calibrating" SLM active coronagraph (Kühn et al. in prep.), based on the phase-shifting Zernike wavefront sensor.[7] LPP: Lyot pupil-plane (as imaged on the CCD after flipping a lens); SFP: scientific focal-plane; RFP: (coronagraphic) reflective focal-plane (SLM location).

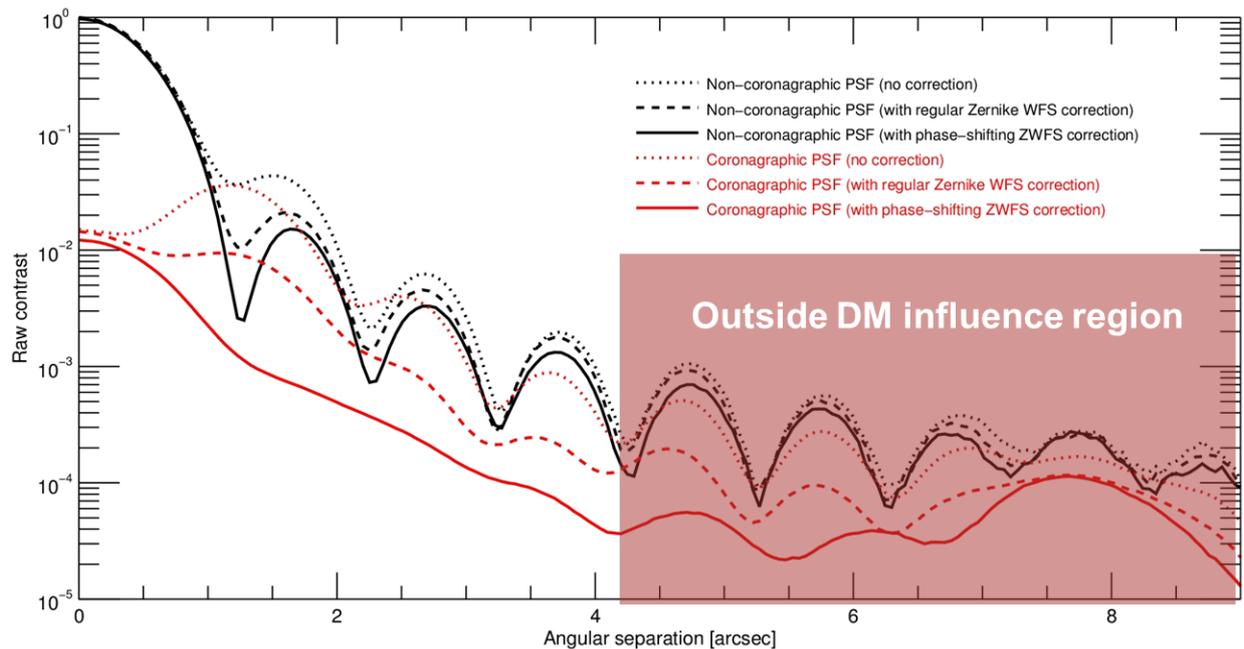

Figure 5. Raw azimuthally-averaged measured contrast curves, with no low-order aberrations correction, with the ZELDA-like single phase-shift Zernike WFS correction,[8] and with the phase-shifting Zernike WFS[7] correction obtained as depicted on Figure 4. The region outside of the DM influence region is indicated.

Although it can be argued that the ZWFS or PS-ZWFS sensing schemes are neither entirely common-path with the science beam train, nor running in real-time during scientific exposure, we note that the Fig.4 SLM-based implementation reduce the gap in various ways: the SLM optical component is the same for coronagraphy or WFS, with no mechanical movement required, the measurements could in principle be quickly interleaved in-between science exposures (assuming a framerate comparable to 30 Hz or faster), and the pupil-plane imaging could be realized in an independent optical arm at a slightly different wavelength. In the latter case, this proposed approach would be free of any mechanical actuation. Finally, Fig.5 presents comparative raw contrast curves before and after correction.

### 3.2 Aperture-optimized FPMs for real telescope pupils

It is common knowledge that unobscured non-ideal telescope pupils represent an imperfect case for coronagraphy, where an off-axis telescope arrangement would be largely preferred for most HCI applications. Concretely, secondary mirror central obscuration, its support structure, and segmented primary mirrors, are responsible for starlight leakage and PSF artifacts in the final image, both affecting achievable contrast and inner-working angle (IWA).[6] Recently, and particularly in the face of upcoming increasingly challenging telescope pupils (WFIRST-AFTA, all ELTs), the HCI community has made large headways to mitigate these issues, relying on a variety of fixed apodizer solutions (amplitude and/or phase), sometimes in combination with multiple optical stages. Nevertheless, the foreseeable slowly-evolving nature of those pupil merit functions, in particular for segmented ELTs primary mirrors, calls for some kind of active solutions. This could be done in a similar fashion as recently achieved to correct slowly-evolving NCPAs for the WFIRST-AFTA coronagraph, using one (or a pair) of DM(s) as active pupil-plane coronagraph (or apodizer),[15] but we also investigated whether active FPMs could play a – possibly complementary – contribution into addressing this challenge. In this vein, we followed up on the work of Ruane et al.,[31] who proposed to use a Gerchberg-Saxton (GS) iterative algorithm to derive optimized FPM solutions for non-ideal telescope pupils, in particular centrally-obscured profiles. As shown in Kühn et al. 2017,[29] multiple FPM solutions to address VLT-like centrally obscured pupils, and even the "spiders" support structure, can be obtained and successfully implemented with a reflective LCOS SLM panel (see examples on Fig.6a). However, the higher rejection generally comes at a cost of throughput, due to locally large phase variations on the FPM, which rarely exceeds 0.1 and exhibit important spatial variations, making the approach only advantageous within the first ~2 $\lambda/D$ from the star.[29] In spite of these drawbacks, the approach holds interesting promises to address defective or non-uniform segmented pupil (see LUVOIR examples on Fig.6b) in the long run, and we note that less aggressive FPM solutions – in terms of throughput - to address centrally-obscured pupils have been recently proposed,[32] motivating further works in the near-future.

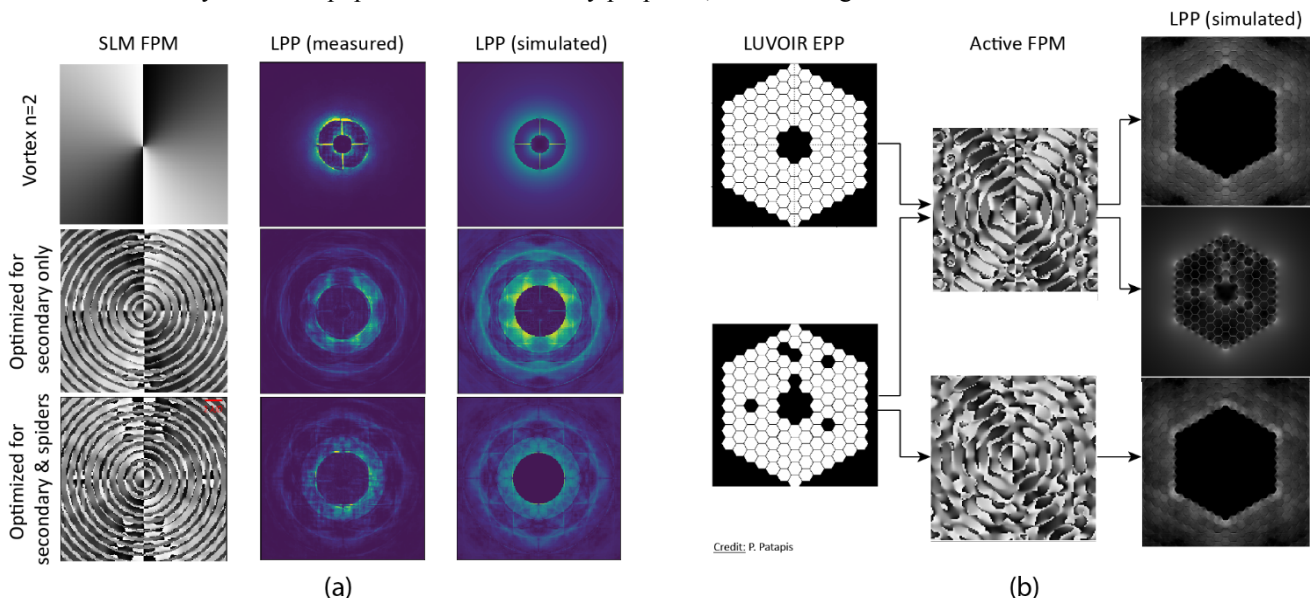

Figure 6. (a) Previously obtained lab results[29] with aperture-optimized focal-plane phase masks (FPMs) programmed onto an SLM active panel, here for the case of a Palomar-shaped entrance pupil-plane (EPP). (b) Potential of active FPMs for segmented pupils presenting missing actuators (e.g. LUVOIR here), where a new ideal FPM can be iteratively retrieved to perfectly diffract starlight outside the downstream Lyot pupil-plane (EPP).

## 3.3 Coronagraphic imaging of multiple stars systems

Binary stars, or stellar systems of higher multiple, represents about half of the stars in our own galaxy. Yet those are frequently excluded for HCI survey, as they prove difficult to observe, and particularly when using focal-plane coronagraphs. This is particularly the case for companions with similar brightness (within 2 stellar magnitudes of the primary), and at close angular separation (less than 0''.5) from the primary stellar component. Less challenging configurations can often be observed using ADI or binary-differential imaging (BDI) techniques. We can however expect the number of tightly resolved binary stars to keep increasing in the era of the ELTs, by the virtue of their improved angular resolution, further stressing the need to develop dedicated technical solutions. In this regard, we recently explored how active SLM-based FPM coronagraphs could deal with this issue, by generating customized optimized FPM phase patterns for a particular on-sky binary configuration (in terms of brightness ration and angular separation), and this for a variety of initial FPM solutions (see Fig.7).[23] As detailed in Kühn et al. 2016,[23] we were able to achieve coronagraphic nulls in excess of $2·10^{-2}$, and this for equal brightness stellar components with angular separation of less than 2 $\lambda/D$. An attractive feature of this approach is the foreseen compatibility with ADI-style pupil-tracking observing, as the SLM panel can be refreshed at video-rate, hence making it possible to "rotate" the phase pattern to "follow" the secondary (or triple, etc.) star(s).

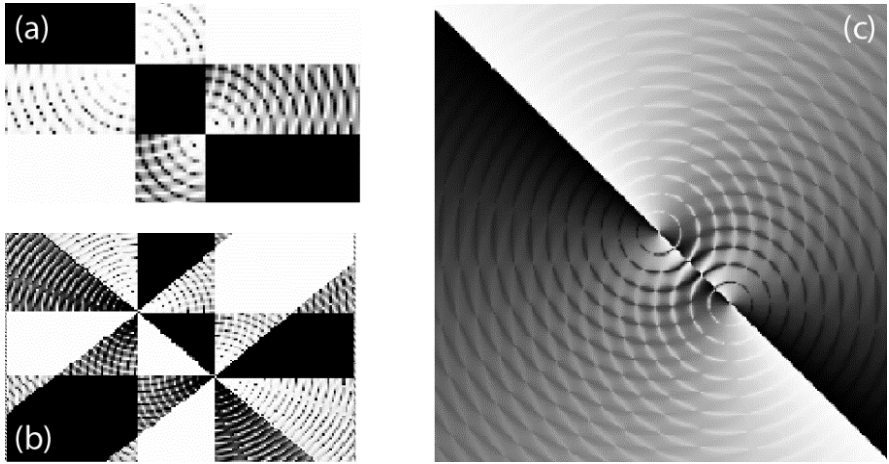

Figure 7. Examples of custom-designed FPM phase patterns optimized for a particular binary stars configuration on-sky, for (a) a FQPM coronagraph, (b) an eight-octants phase mask, and (c) a vortex of topographic charge n=2. Those masks could be rotated at the video-framerate of the SLM, for example to compensate for sky rotation during an ADI observation.

## 3.4 Coherent Differential Imaging (CDI) by local phase diversity in the coronagraphic focal-plane

Taking advantage of the mutually temporally coherent nature of the NCPAs speckles with respect to the stellar coronagraphic PSF, while any bona-fide off-axis astrophysical source would retain complete incoherence, coherent differential imaging (CDI) concepts have long been theorized to be able to provide an extra one to two orders of magnitudes of contrast with next to no extra hardware overhead.[12] In practice, however, the first actual successful implementations are only burgeoning now, usually limited to some blind speckles nulling iterative wavefront retrieval, whereas somewhat faster solutions would be desirable in the future, to make the best use of the "freezing" of atmospheric residual speckles that could be achieved with low-noise high-speed detectors like micro-kinetics inductance arrays (MKIDS).[33] To this end, we recently started exploring the CDI capabilities of a SLM-based active FPM, by using the SLM to locally phase-shift an individual, or a series of, speckle(s) in the coronagraphic focal-plane, hence performing phase diversity at a known "carrier" modulation frequency (see Fig.8). We then concatenated the acquired data cube, and performed an analysis in the temporal Fourier domain (see Fig.8), by building a simple empirical estimator for the coherence, as the ratio of the power spectrum signal at the carrier frequency versus the DC component (Fig.8). In parallel, we used a different laser source to feed the 2$^{nd}$ port of our optical setup (Fig.1a) with a weaker signal, effectively adding a ~$10^{-4}$ incoherent "fake planet" at a separation of ~4.5 $\lambda/D$ from the central star. Figure 9 shows the results of the empirical coherent – and incoherent (the inverse) – estimator 2-D mappings, where the lack of coherence (or excess of incoherence) at the location of the "fake planet" is clearly evidenced. Promisingly, Fig.9 also displays similarly encouraging results using much larger, but less numerous, ring-shaped modulation regions, hinting at possibly more time-effective and practical CDI detection schemes. Finally, we note that the preliminary results of Fig.9 are still very early material, and that much needs to done to optimize the modulation process (number of phase shifts, here 8, size and overlap of the modulation regions, …), estimate the S/N gain in function of angular separation, and how to best converge to a useful final image or coherence metric.

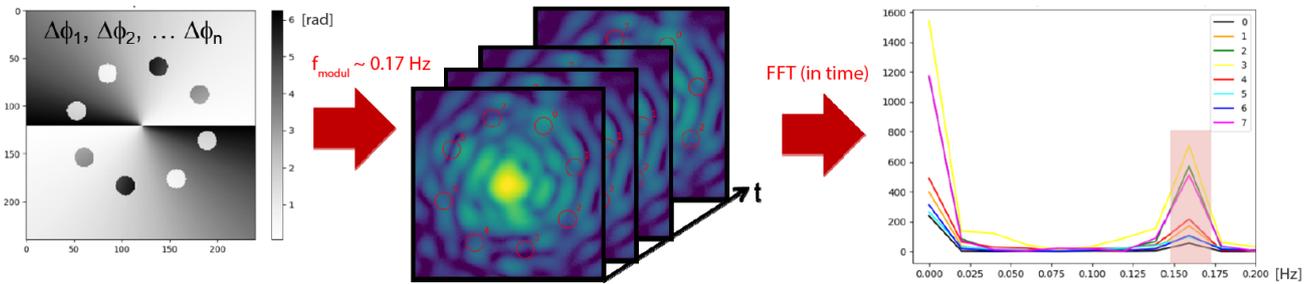

Figure 8. Principle of focal-plane phase diversity with a SLM active coronagraph to achieve CDI. (Left) Various regions-of-interests (ROIs) are phase-modulated in time at given frequency (here 8 ROIs undergo 8 phase shifts at 0.17 Hz), then (Middle) the temporal data stack is concatenated, and (Right) the Fourier transform (power spectrum) in time is taken to isolate the DC and carrier-frequency components for each ROI (here the incoherent ~$10^{-4}$ "planet" correspond to ROI #0).

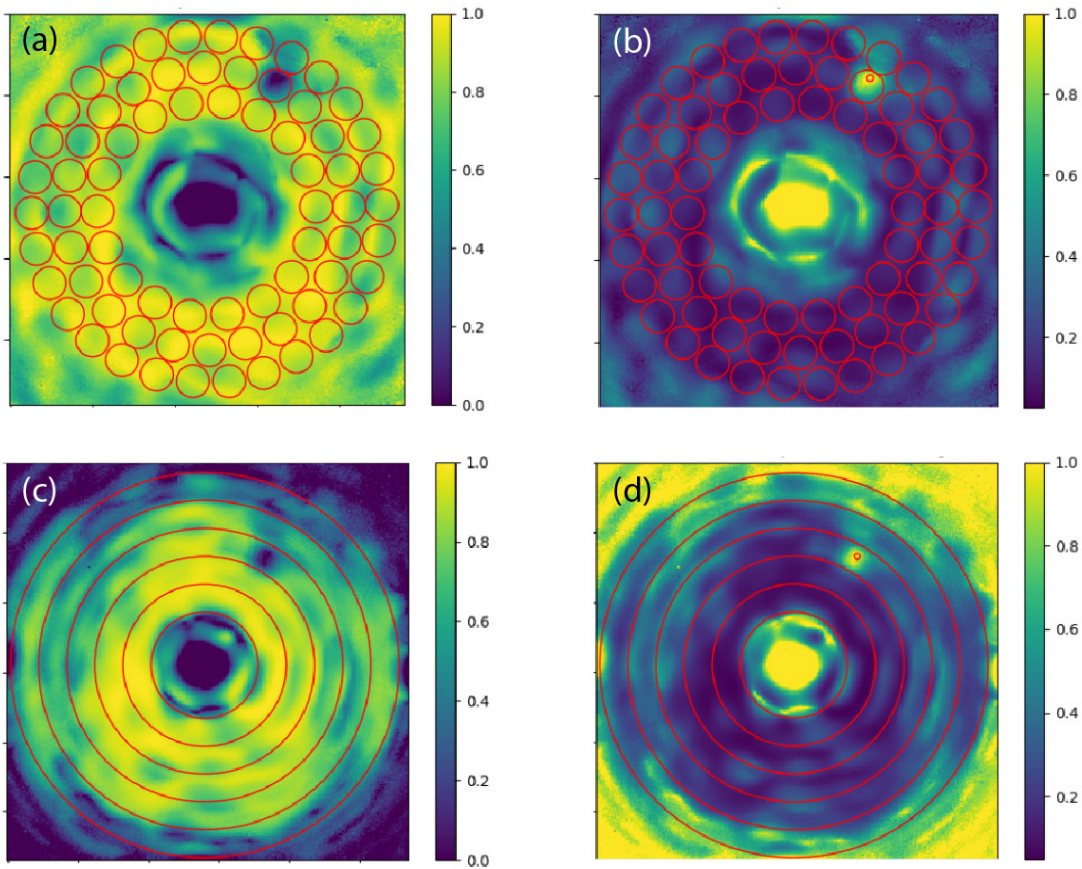

Figure 9. Preliminary CDI results when processing the phase diversity data according to Fig.8, and estimating the (a) coherent or (b) incoherent fields as simple ratios of power spectrum DC and carrier frequency components. (c,d) Same but for larger 1 λ/D-wide ring-shaped modulation ROIs.

# 4. CONCLUSIONS AND PERSPECTIVES

Given the current and upcoming challenges in direct imaging instrumentation, the need to integrate additional adaptive or active optics components in future HCI platforms is a likely scenario, particularly to be able to re-configure the instrument to deal with changing pupil conditions, in particular a segmented primary mirror apertures with non-static merit function. This would represent a natural evolution of nowadays instruments of the likes of GPI, SPHERE and SCExAO, which have shown an increasing reliance on AO and WFS sub-systems at various locations along the beam train, as compared to first-generation instruments. Further, the near-term availability of novel NCPAs real-time WFS techniques, and scientific FPAs with sub-ms temporal resolution, is obviously calling for some parallel integration of dynamic wavefront and beam shaping control technologies along the science optical path downstream of the facility AO system, ideally operating at similar temporal bandwidth. That should notably include some kind of active coronagraphic optics, and even more so as the finality of a HCI instrument is to reach the deepest possible contrast (i.e. optimal starlight rejection), not to achieve a perfect wavefront correction. This subtle difference is often ignored, as static coronagraphs are inherently performing better under ideal wavefront conditions, but it might come into play in a dynamic adaptive scenario. Finally, another promising approach to differentiate between stellar residuals and off-axis astrophysical signal is coherent differential imaging (CDI), which relies on the coherence properties of the starlight to discriminate the stellar PSF residuals from actual bona-fide incoherent astrophysical sources. The CDI scheme has long been theorized to be able to potentially improve contrast floor of existing instruments by at least two orders of magnitudes, yet practical on-sky implementations are only emerging now (Bottom et al. 2017). In this regard, a dynamic system operating at milliseconds timescale might be able to use the temporal domain to probe the electric field for coherence, possibly paving the way to synchronous demodulation techniques in the wake what we attempted here (§3.4), effectively filtering the coherent noise out from the temporal data stack, thus improve signal-to-noise (S/N) ratio on incoherent sources.

In the wake of our early results, whether LCOS-SLM is the right technology for active coronagraphy in astronomical high-contrast imaging remains to be seen, due to its few inherent potential flaws. In this regard, the SLM technology limitations regarding cryogenic operations (for mid-IR wavelengths), and use of linearly-polarized light for phase-only operations, are unlikely to change in a foreseeable future. The limited coronagraphic attenuation and chromatic nature of SLMs may actually end up being less of an issue, as wavefront control is likely to remain the first performance bottleneck for a while, particularly on ground-based observatories, while various workaround solutions addressing those limitations remain to be investigated.[27] Additionally, it is worth underlining that such SLM-based active coronagraphs can be instantaneously made "perfect" for a given wavelength, and this by a pure software operation, potentially enabling new routes for broadband coronagraphy when photons scarcity is a non-issue: for example one could synchronize an IFS spaxel readout with a wavelength-scanned coronagraphic SLM phase pattern. More generally, most of the investigations into the modalities and capabilities of SLM-based active coronagraphs will maintain relevance when dealing with alternative optical modulation technology (e.g. DMDs, MEMS-DMs …), which to our sense should fully motivate the community to look into this new approach of doing coronagraphy.

# ACKNOWLDGEMENTS

This work was performed at the Institute of Particle Physics and Astrophysics of ETH Zurich, with the "Digital Adaptive Coronagraphy Project" and JK being funded by the Swiss National Science Foundation (SNSF) through Ambizione grant #PZ00P2_154800. We hereby acknowledge all the participants of the "active coronagraphy" session of the Leiden 2017 "Optimal Optical Coronagraphs" workshop for the inputs and helpful contributions leading to the discussion in §2.1, namely Dr. O. Absil, Dr. P. Baudoz, Dr. A. Carlotti, Dr. M. N'Diaye, Dr. R. Galicher, Prof. Dr. O. Guyon, Dr. J. Mazoyer, and Dr. E. Por. We also thank Prof. H. M. Schmid for the continuous logistics and travel funding support throughout the project.